\documentclass[authoryear,10pt,5p,times]{elsarticle}

\makeatletter
\def\ps@pprintTitle{%
 \let\@oddhead\@empty
 \let\@evenhead\@empty
 \def\@oddfoot{\centerline{\thepage}}%
 \let\@evenfoot\@oddfoot}
\makeatother

\usepackage{graphicx}

\begin{document}
\begin{frontmatter}
\title{Estimation of error on the cross-correlation, phase and time lag  between evenly sampled light curves}


\author[rm]{Ranjeev Misra}
\ead{rmisra@iucaa.in}
\address[rm]{IUCAA, Post Bag 4, Ganeshkhind, Pune-411007, India}

\author[ab]{Archana Bora}
\ead{abora.80@gmail.com}
\address[ab]{Dept. of Applied Sciences, Gauhati University Institute of Science and Technology, Guwahati 781014, India}

\author[gd]{Gulab Dewangan}
\ead{gulabd@iucaa.in}
\address[gd]{IUCAA, Post Bag 4, Ganeshkhind, Pune-411007, India}

\begin{abstract}
Temporal analysis of radiation from Astrophysical sources like
  Active Galactic Nuclei, X-ray Binaries and Gamma-ray bursts provide
  information on the geometry and sizes of the emitting regions.
  Establishing that two light-curves in different energy
  bands are correlated and measuring the phase and time-lag between
  them is an important and frequently used temporal diagnostic.
   In this work, we have presented expressions for estimate of the errors on the cross-correlation,
phase and time-lag between two light-curves
 and the same have been tested using simulations.
  Earlier estimates depended upon numerically expensive
  simulations or on dividing the light-curves in large number of
  segments to find the variance. 
  The estimates
  presented here allow for analysis of light-curves with relatively
  small ($\sim 1000$) number of points, as well as to obtain
  information on the longest time-scales available. 
  For testing the analytical expressions light-curves have been simulated from
  both white and $1/f$ stochastic processes with measurement
  errors. As a demonstration, we also apply this technique to the {\it
  XMM-Newton} light-curves of the Active Galactic Nucleus, Akn~564.

\end{abstract}


\end{frontmatter}

\section{Introduction}
\label{intro}

Establishing that two light-curves, measured in different energy
bands, are correlated with each other is an important temporal
diagnostic for various kinds of Astrophysical sources, especially for
Active Galactic Nuclei (AGN) and X-ray binaries. The detection and
measurement of the level of correlation constrains the 
radiative processes active in the source and can be used to validate
(or rule out) models based on spectral analysis.  Phase and time-lags
detected for correlated light-curves can provide further insight into
the geometry and size of the emitting region. Often in these
applications, the light curves available for analysis are of short
duration and have measurement errors. The true temporal behaviour of a
source can only be established if there are robust estimates of the
errors on the cross-correlation, phase and time-lags.

It is important to emphasize that a cross-correlation analysis between
two finite length light-curves will not provide an accurate measure of
the correlation between them, even in the absence of measurement
errors. Intrinsic stochastic fluctuations in the light curves will
induce an error on the cross-correlation measured.  An estimate of the
significance and error of the cross-correlation detected, should take
into account both, measurement errors as well as statistical
fluctuations.

A standard method to estimate the error on the cross-correlation
involves dividing the light curves into several equal segments and
finding the cross-correlation for each. Then the net cross-correlation
is given by the average of the different segments and the variance is
quoted as an error. For example, this technique is implemented by the
function ``crosscor'' of the high energy astrophysics software {\it
HEASOFT\footnote{http://heasarc.gsfc.nasa.gov/docs/software/lheasoft/}}. The
method is reliable only if the light curves can be divided into a
large number of segments ($>> 10$) and each segment is sufficiently
long and not dominated by measurement errors. The temporal behaviour
of many astrophysical systems depends on the time-scales of the
analysis and hence by using this method, one loses information on the
behaviour of the system on time-scales comparable to the length of the
original data. In AGN, the time scale involved is long comparable to
the length of observation in many cases, hence it is not practical to
divide the light curve in segments. Moreover, there does not seem to
be any established way by which this method can be extended to get an
estimate of the time-lag between the light curves and its error.

These deficiencies can be overcome by using a Monte Carlo technique
where one simulates a large number of pairs of light curves having the
same {\it assumed} temporal properties and with the same measurement
errors as the original pair. The results of the original pair can be
compared with the simulated ones to ascertain the confidence level of
the cross-correlation and time-lag. The simulated light curves should
take into account the stochastic fluctuations of the light curves and
not just the measurements errors. Indeed, when the light curve is
sampled unevenly and with measurement errors changing in time, the
Monte Carlo technique may be the only way to obtain reliable estimates
\cite[for e.g.][]{peterson1998uncertainties}. Monte Carlo technique is numerically
expensive and hence are not practical for analysis of a large sets of
data. More importantly, the results depend on the subjectivity of the
assumed temporal properties of the system. For example, to ascertain
the errors on an observed cross-correlation and time-lag value, the
simulations are generally done with the assumption that these are the
true intrinsic values. Similar assumptions have to be made on the
shape of the power spectra of the light curves.

As pointed out and discussed extensively by \cite{welsh1999reliability}, an
analytical estimate of the variance on cross-correlation is not
straight forward. In the literature, there is an analytical estimate
for the cross-correlation known as Bartlett's equation \cite{bartlett1955introduction}
which is not often used in Astronomical contexts. This method is
available in the ``crosscorrelation'' function in the IMSL numerical
libraries\footnote{http://www.vni.com}. The error is accurate only
when the complete knowledge of the cross-correlation and
auto-correlation functions are available.  Its effectiveness for short
duration light curves is uncertain. Moreover, this error estimate does
not naturally translate into error estimates for the phase and time
lag between the light curves.

Complete information regarding the temporal relation between two light
curves can be obtained by computing the coherence and time-lag as a
function of Fourier frequency. A detailed description of the technique
as well as physical interpretation is given by \cite{nowak1999rossi}. The two
light curves are divided into many segments and for each segment a
Fourier transform is undertaken and coherence and phase lag as a
function of frequency is estimated. For the different segments, the
coherence and phase lags are averaged and their errors can be
estimated analytically.  Such detailed information can only be
obtained for long light curves which can be split into several
segments.  In the absence of such rich data, statistically significant
results can be obtained by averaging over Fourier frequencies. Indeed,
from this view point the cross-correlation, is in some sense, the
average of the coherence over all frequencies. However, computing the
error on the cross-correlation using the error estimates for the
coherence is not straight forward. First, the averaging has to be
appropriately weighted by the power in each frequency bin.  Secondly,
the error estimate for the coherence is reliable only if the error
itself is small, which is the case when many segments are averaged and
not necessarily true for the coherence at a single frequency bin
obtained from a single segment.

In this work, we present an 
expression for the
cross-correlation between two evenly sampled light curves. The error estimate is
based on the Fourier transforms of the light curves then averaging over different
frequency modes. 
\textsection 2 presents the expression 
and the same has been verified by simulations with and without
measurement errors.  \textsection 3 highlights the difficulties in estimating a
time-lag and its error using the standard method of finding the peak
of the cross-correlation function. The cross-Correlation phasor is
introduced in \textsection 4 which leads to an estimate of the phase lag between
the light curves.  In the same section, a technique is introduced by
which one can measure the time lag and its error. In this method the
time-lag measured can be even smaller than the sampling time bin of
the light curves. The complete fully self contained algorithm is
presented in \textsection 5 for easy reference. As an example, in \textsection 6, the
technique is applied to the {\it XMM-Newton} light curves of the
highly variable and well studied AGN, Akn~564. In \textsection 7, the summary
and discussion includes a list of important assumptions on which the
technique is based and provides examples when the assumptions may not
be valid.

\section{ANALYTICAL ERROR ESTIMATE OF \\CROSS-CORRELATION}
\label{aeecr}

\subsection{Light curves without measurement errors}

 We first consider an idealised case of two light curves, $X$ and $Y$, without
measurement errors. The two light curves are of length N, which have recorded the count rates
in $j=0,1,2,...,N-1$ discrete equally spaced time intervals, $\Delta t$. The mean
is subtracted from each of them. Further it is assumed that they are partially
linearly dependent on each other by  {\textit{A}}, such that we have,
\begin{eqnarray}
  X & = & x_j \\
  Y & = & z_j + Ax_j
\end{eqnarray}
where $x_j$ and $z_j$ are time-series produced by two independent
stochastic processes. Each time series can be conveniently represented, in frequency domain $k$
by its discrete Fourier transform, $\tilde{X}_{k}$, defined as
\begin{equation}
  \tilde{X}_{k} = \sum_{j=0}^{N-1} {X}_{j} \exp{(2\pi ijk/N)}
\end{equation}
and a power spectrum is estimated as $P_{Xk} \equiv (2/N) |\tilde{X}_k |^2$.
Here the normalisation constant of the power spectrum is used as suggested by
Leachy et al \cite{leahy1983searches}. For a stationary system, the ensemble average (i.e. average
of an infinite number of realisations) of the power, $<P_{Xk}>$, is a
characteristic of the stochastic process. A power derived from a
single time series, $P_{Xk}$ is only an estimator of its value. In
particular the real and imaginary parts of $\tilde{X_k}$ varying
independently can be derived from two independent Gaussian
distributions \cite{timmer1995generating}. The standard deviation of $P_{Xk}$ from 
$<P_{Xk}>$ is roughly equal to $<P_{Xk}>$  i.e the power estimate from a single
light curve has nearly 100\% sampling variation. The variance $\sigma^2_X \equiv \sum
P_{Xk}$ is again an estimate of the ensemble averaged variance
$<\sigma^2_X> = \sum <P_{Xk}>$ \cite{van1989fourier} where $k=-N/2,....,N/2-1$ and $k\neq 0$.

One can define the cross-correlation estimate of the two time series
as
\begin{equation}
  C_{XY}  = \frac{c_{XY}}{\sqrt{\sigma^2_{X} \sigma^2_Y}}
\end{equation}
where
\begin{equation}
  c_{XY}  = = \frac{1}{N}\sum_{j=0}^{N-1} X_j Y_j = \frac{1}{N^2}\sum_{k=-N/2}^{N/2-1} {\tilde X}_k {\tilde Y}_k^*
  \label{smallcxy}
\end{equation}
Here ${\tilde X}_k$ and ${\tilde Y}_k$ are Discrete Fourier transforms
of $X_j$ and $Y_j$ respectively and
\begin{eqnarray}
  \sigma^2_X & = & \frac{1}{N}\sum_{j=0}^{N-1} X_j^2  =  \frac{1}{N^2}\sum_{k=-N/2}^{N/2-1} |{\tilde X}_k|^2 \nonumber \\
  \sigma^2_Y & = & \frac{1}{N}\sum_{j=0}^{N-1} Y_j^2  =  \frac{1}{N^2}\sum_{k=-N/2}^{N/2-1} |{\tilde Y}_k|^2 
  \label{sigmaxy}
\end{eqnarray}
Their ensemble averages are {$<c_{XY}> = A <\sigma_x^2>$},
$<\sigma_X^2> = <\sigma_x^2>$ and $<\sigma_Y^2> = <\sigma^2_z> +
A^2<\sigma^2_x>$. It is to be noted that a normalisation factor of $\frac{1}{N}$, due to the DFT process,
has been incorporated in frequency domain term of Eqn (\ref{sigmaxy} ) $C_{XY}$ has the useful property that its ensemble
average
\begin{equation}
  <C_{XY}> = \frac{A <\sigma_x^2> }{\sqrt{<\sigma_x^2>( <\sigma^2_z> + A^2<\sigma^2_x>)}}
\end{equation}
is zero if the two light series are uncorrelated (i.e. $A = 0$) and
$\pm 1$ if they are completely correlated (i.e. when { $<\sigma^2_z> =
0$}). 
However, in absence of any a priori information about the stochastic process, 
quantities need to be estimated using the measured values only.
Thus, $C_{XY}$ is also an estimate of $<C_{XY}>$ and its accuracy needs
to be quantified.

To ascertain whether there is a detectable correlation between the two
light-curves (i.e. $|C_{XY}| > 0$) it is first necessary to show that,
at some confidence level, $|c_{XY}| > 0$. One can define a null
hypothesis sigma level $\sigma_{NH} = |C_{XY}|/\Delta C_{XY}^\prime$
and fix a criterion (a prudent one being $\sigma_{NH} > 3$) to
ascertain whether any correlation has been detected. It is important
to note that only if the criterion is satisfied should one proceed to
estimate the degree of cross-correlation $C_{XY}$ otherwise any such
attempt will not only be incorrect but also meaningless.

If $c_{XY}$ is uncorrelated with $\sqrt{\sigma^2_{X} \sigma^2_Y}$,
then
\begin{equation}
  \Delta C_{XY} = \frac{ \Delta c_{XY}}{\sqrt{\sigma^2_{X} \sigma^2_Y}}
\end{equation}
where the variation in $\sqrt{\sigma^2_{X} \sigma^2_Y}$ has been
neglected.  However, as discussed extensively by \cite{welsh1999reliability}, $
\sqrt{\sigma^2_{X} \sigma^2_Y}$ is, in general, correlated with
$c_{XY}$. In particular, $\Delta c_{XY}$ depends on the variation of
$\sigma^2_X$ through the term $A \sigma^2_X$.
 
A possible solution is to define a transformation, $P(C_{XY})$ whose
terms are not correlated (or at least not so correlated).  Then
estimate the expected variation for that function, $\Delta P$ and use
that to obtain an estimate for $\Delta C_{XY}$.  Below we describe
such a transformation and subsequently test the results obtained from
simulations.  The transformation choosen for the analysis is
\begin{equation}
  P = \frac{c_{XY}^2}{\sigma^2_{X} \sigma^2_Y - c_{XY}^2} = \frac{C_{XY}^2}{1 - C_{XY}^2}
\end{equation}
where the subtraction of $ c_{XY}^2$ in the denominator may make it
nearly independent of the numerator. 
The average deviation of $P$ can
be estimated to be,
\begin{equation}
  \Delta P = \frac{2 <c_{XY}> \Delta c_{XY}}{<{\sigma^2_{X} \sigma^2_Y} - c_{XY}^2>}
 \label{eqnp1}
\end{equation}
where the variation of the denominator has been neglected. $\Delta P$
is related to $\Delta C_{XY}$ by
\begin{equation}
  \Delta P \sim <(\frac{dP}{dC_{XY}})>\Delta C_{XY} = <\frac{2 C_{XY} }{(1 - C_{XY}^2)^2}> \Delta C_{XY}
  \label{eqnp2}
\end{equation}

giving us
\begin{equation}
  \Delta C_{XY} = \frac{1 - <C_{XY}>^2}{\sqrt{<\sigma^2_{X}> <\sigma^2_Y>}} \Delta c_{XY}
  \label{dcxy}
\end{equation}

Naturally, $\Delta C_{XY}$ depends on ensemble averaged quantities
which characterise the stochastic processes that have produced the
light curves. Typically, one does not have {\it a priori} information
of the stochastic processes and the ensemble averaged quantities need
to be estimated from the light curves. Thus the best estimate, $\Delta C_{XY}^\prime$, of
$\Delta C_{XY}$ can be obtained by replacing these ensemble averaged
quantities with the measured ones. Also, we have 
$ (\Delta c_{XY})^2= <c_{XY}^2>-<c_{XY}>^2=\frac{1}{N^4}\sum_{k=-N/2}^{N/2-1} <|{\tilde X}_k|^2><|{\tilde Y}_k|^2> 
$

Hence
\begin{equation}
  \Delta C_{XY}^\prime = \frac{1 - C_{XY}^2}{N^2 \sqrt{\sigma^2_{X}\sigma^2_Y}} \sqrt{\sum_{k=-N/2}^{N/2-1}|{\tilde X}_k|^2|{\tilde Y}_k|^2}
  \label{dcxy_p}
\end{equation}
For practical situations $\Delta C_{XY}^\prime$ can be used as an
estimate for the error on $C_{XY}$.

\subsection{Comparison with results from simulations}

We generated 200 independent light-curves using the method described
by \cite{timmer1995generating}. \cite{vaughan2003characterizing} discuss the different methods to
generate stochastic light curves and give arguments for favouring the
one prescribed by \cite{timmer1995generating}. The intrinsic power spectrum of the
stochastic process was assumed to be a power-law i.e. $P (f) \propto
f^{-\alpha}$. The light-curves were generated of length $8N$ and
rebinned to a length of $N$, to avoid aliasing effects. From these 200
light curves, $19900$ pairs of the light curves were generated which
obey,
\begin{eqnarray}
  X_j & = & x_j \\
  Y_j & = & z_j + Ax_j
\end{eqnarray}
where $x_j$ and $z_j$ are two different simulated light-curves with ${<\sigma^2_x>}={<\sigma^2_z>} = 1$.  The
cross-correlation, $C_{XY}$ was computed for each pair. For $N =
1024$, $\alpha = 0$ and for three different values of $A = 0,1$ and
$5$, the histograms of the $C_{XY}$ are plotted in the top panel of
Figure 1. These histograms, $H_j$ are normalised such that their
summation $\sum H_j \delta = 1$ where $\delta$ is the bin size. They
are compared with a normalised Gaussian distribution with a centroid
value equal to the expected averaged cross-correlation of
\begin{equation}
  <C_{XY}> = \frac{A}{\sqrt{1 + A^2}}
\end{equation}
\begin{figure}[t]
  \begin{center}
    {\includegraphics[width=\linewidth]{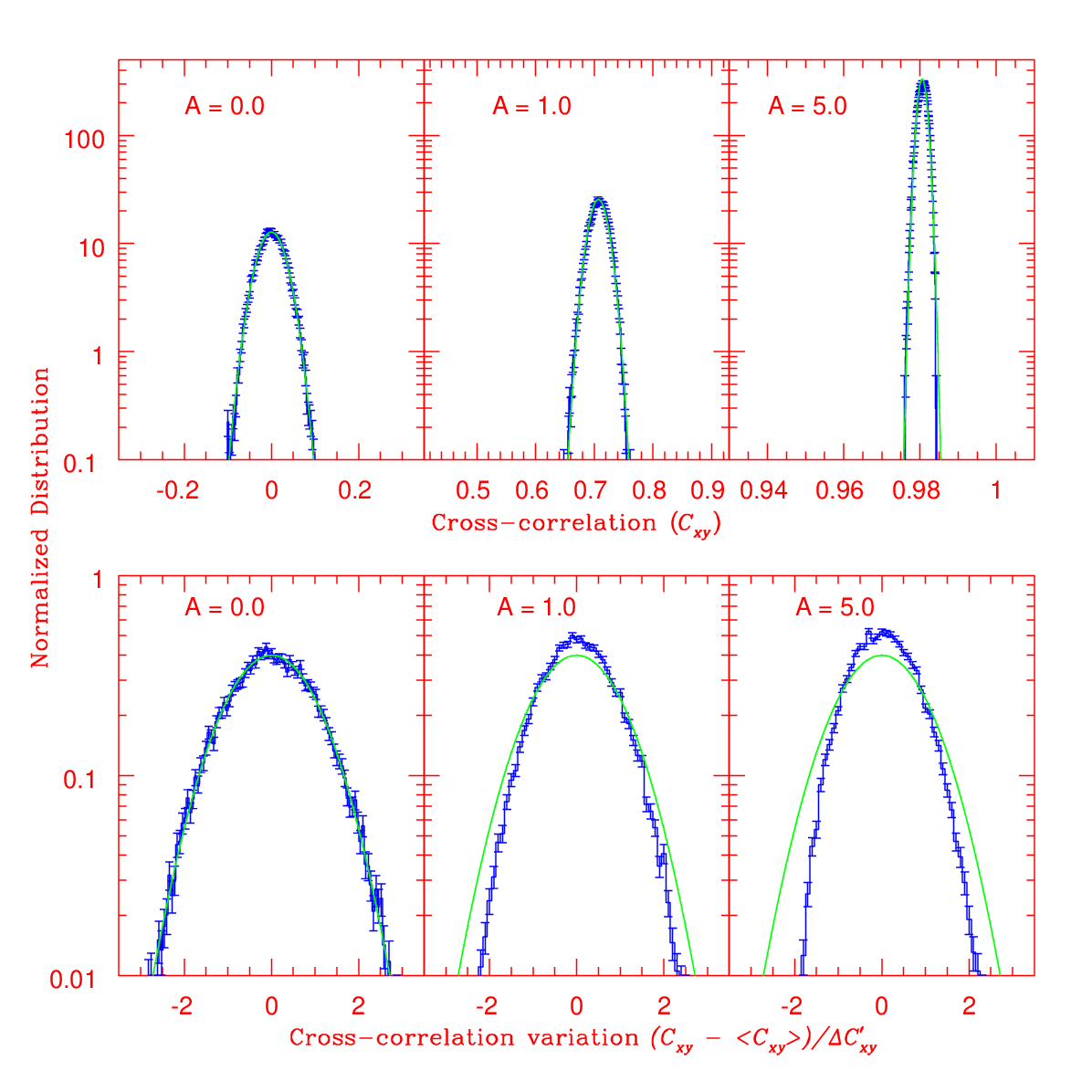}}
  \end{center}
    \caption{Comparison of simulation with analytical results. 19900
  pairs of light-curves of length $N = 1024$ were created for $X_j =
  x_j$ and $Y_j = z_j + Ax_j$. $x_j$ and $z_j$ are independent
  time-series generated from a stochastic white noise process
  (i.e. power spectrum index $\alpha = 0$). Top Panel compares the
  normalised histogram of $C_{XY}$ with a Gaussian with centroid at
  the expected $<C_{XY}>$ and width given by $\sigma = \Delta C_{xy}$
  (Eqn \ref{dcxy}). Bottom panel shows the histogram of the
  cross-correlation variation $(C_{XY} - <C_{XY}>)/\Delta
  C_{XY}^\prime$. If $\Delta C_{XY}^\prime$ (Eqn \ref{dcxy_p}), which
  is estimated from the pair of light-curves, is a true measure of the
  variation, then the distribution should be a zero centred Gaussian
  with $\sigma = 1$ (solid line).  }
\end{figure}

\begin{figure}
  \begin{center}
    {\includegraphics[width=\linewidth]{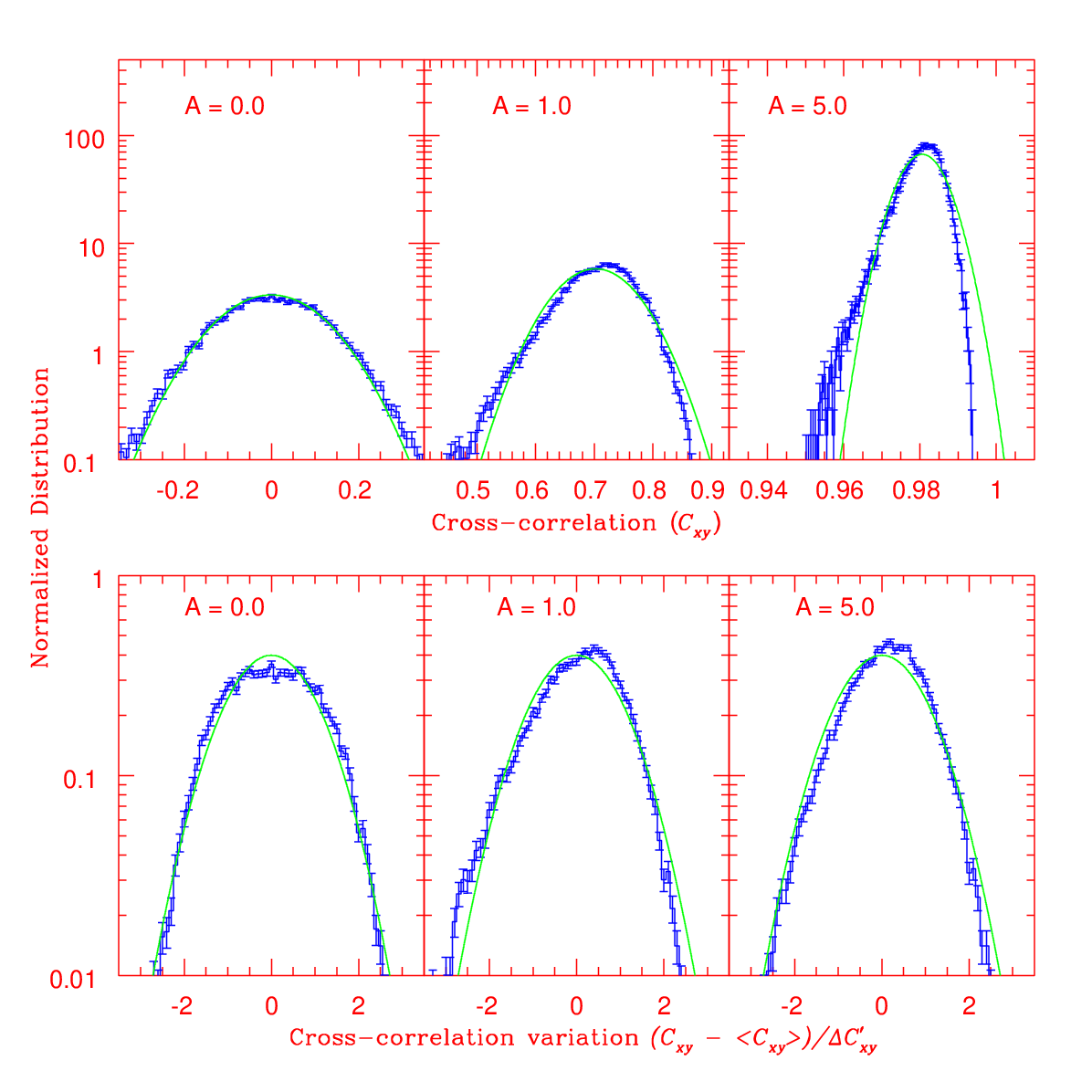}}
  \end{center}
  \caption{ Same as Figure 1, except that $x_j$ and $z_j$ are
  independent time-series generated from a stochastic $1/f$ noise
  process (i.e. power spectrum index $\alpha = 1$). The width of the
  distributions are broader than for white noise.}
\end{figure}
and with width $\sigma$ equal to $\Delta C_{XY}$ computed using Eqn
(\ref{dcxy}). As can be seen, the normalised Gaussian distributions
describe well the simulated results which validates the method and
assumptions used to estimate $\Delta C_{XY}$ in the previous
subsection. However, in practical situations one has to use the
approximation $\Delta C_{XY}^\prime$ to estimate the variance which in
general will vary for each pair of light curves.  In the bottom panel
of Figure 1, we plot histograms of deviation of $C_{XY}$ from the
average $<C_{XY}>$ normalised by the estimated deviation $\Delta
C_{XY}^\prime$ i.e. $(C_{XY} - <C_{XY}>)/\Delta C_{XY}^\prime$. If
$\Delta C_{XY}^\prime$ is an accurate measure of the variation of
$C_{XY}$ then the distribution of the normalised variation should be a
zero centred Gaussian with $\sigma = 1$. The plot verifies this
prediction by comparing the distribution with such a Gaussian
shape. The distributions agree well with each other except for large
$A$, where the Gaussian distribution is slightly broader than the
simulation results. This implies that the $\Delta C_{XY}^\prime$ is a
slight overestimation of the true deviation for large values of $A$.

Figure 2 shows the same comparison as Figure 1, but for the case when
the power-law index $\alpha = 1$. Qualitatively the comparison between
the expected and obtained distribution are similar to the $\alpha = 0$
case, except for some quantitative differences. For the same length of
the light curves, $\Delta C_{XY}$ is larger for $\alpha = 1$.  Since
$C_{XY}$ is by definition constrained to be less than unity, the
distribution differs from the symmetric Gaussian shape for large
$A$. The bottom panel shows that $\Delta C_{XY}^\prime$ is a better
representation of the variation than it was for $\alpha = 0$.

For white noise (i.e. $\alpha = 0$), the dependence of $\Delta C_{XY}$
on the length of the light-curves is $\propto 1/\sqrt{N}$, while for
$\alpha = 1$ the dependence is weaker $\sim 1./\hbox{log}N$ for large
$N$ \cite{chatfield2016analysis, keshner19821}.  The original light-curves may be divided into $M$ parts, and
cross-correlations of each may be averaged. For $\alpha = 0$ this will
not lead to any change in the accuracy with the final $\Delta C_{XY}$
being nearly the same.  However, for $\alpha = 1$, $\Delta C_{XY}
\propto 1/(\sqrt{M}\hbox{log}(N/M))$ which would give a much better
accuracy than finding the cross-correlation for the whole
light-curve. However, such a cross-correlation will not have
information about the behaviour of the system on timescales
corresponding to duration of the original light curve.


\subsection{Light curves with measurement errors}

We next consider a more realistic case, where the light-curves
have measurement errors. In particular,
\begin{eqnarray}
  X_j & = & x_j + e_{Xj} \nonumber \\
  Y_j & = & z_j + Ax_j + e_{Yj}
  \label{lcerr}
\end{eqnarray}
where $x_j$ and $z_j$ are time-series produced by two independent
stochastic processes as before and $e_{Xj}$ and $e_{Yj}$ are the known
measurement errors for measuring $X_j$ and $Y_j$ respectively. The
cross-correlation is now defined as
\begin{equation}
  C_{XY}  = \frac{c_{XY}}{\sqrt{(\sigma^2_{X} - \sigma^2_{XE}) (\sigma^2_{Y}-\sigma^2_{YE})}}
\end{equation}
where $c_{XY}$ is the same as before (Eqn \ref{smallcxy}) and
$\sigma_{XE}$ and $\sigma_{YE}$ are the rms variation of the measured
errors i.e.
\begin{equation}
  \sigma^2_{XE} = \frac{1}{N}\sum_{j=0}^{N-1} e_{Xj}^2
\end{equation}
and similarly for $\sigma^2_{YE}$.

The expressions for $<c_{XY}>$ and $\Delta c_{XY}$ remain the same as
for the measurement error free case discussed previously and following
the same procedure as before, one can estimate
\begin{equation}
  \Delta C_{XY} = \frac{(1 - <C_{XY}>^2) \Delta c_{XY} }{\sqrt{(<\sigma^2_{X}>-<\sigma^2_{XE}>)(<\sigma^2_Y>-<\sigma^2_{YE}>)}} 
  \label{dcxyerr1}
\end{equation}
analogous to Eqn (\ref{dcxy}). To this error estimate we have to add
the fluctuations of $\sigma_{XE}$ and $\sigma_{YE}$ around their
ensemble averaged values $<\sigma_{XE}>$ and $<\sigma_{YE}>$. Note
that it is these ensemble averaged values $<\sigma_{XE}>$ and
$<\sigma_{YE}>$ that are known {\it a priori} and not $\sigma_{XE}$
and $\sigma_{YE}$.  If the measurement errors are Gaussian white noise
(as is generally the case) then, $\Delta \sigma^2_{X,YE} =
(1/\sqrt{N})\sigma^2_{X,YE}$.  Moreover since the fluctuations are
independent of the true signal, they can be added to $\Delta C_{XY1}$
using standard error propagation techniques.  Thus
\begin{eqnarray}
  (\frac{\Delta C_{XY}}{<C_{XY}>})^2 & = &  (\frac{\Delta C_{XY1}}{<C_{XY}>})^2 + (\frac{\Delta \sigma^2_{XE}}{\sqrt{2} (<\sigma^2_X> - <\sigma^2_{XE}>)})^2 \nonumber \\
  & & + (\frac{\Delta \sigma^2_{YE}}{\sqrt{2} (<\sigma^2_Y> - <\sigma^2_{YE}>)})^2 \nonumber \\
  & = & (\frac{\Delta C_{XY1}}{<C_{XY}>})^2 + (\frac{ <\sigma^2_{XE}>/\sqrt{2N}}{ <\sigma^2_X> - <\sigma^2_{XE}>})^2 \nonumber \\
  && + (\frac{<\sigma^2_{YE}>/\sqrt{2N}}{<\sigma^2_Y>- <\sigma^2_{YE}>})^2\nonumber \\
  && 
  \label{dCxyerr}
\end{eqnarray}
$\Delta C_{XY}$ is in terms of ensemble averaged quantities which have
to be estimated using the light curves. Hence
\begin{eqnarray}
  \Delta C^\prime_{XY} & = & \frac{(1 - C_{XY}^2) }{N^2 \sqrt{(\sigma^2_{X}-<\sigma^2_{XE}>)( \sigma^2_Y-<\sigma^2_{YE}>)}} \nonumber\\
  & & \times \sqrt{\sum_{k=-N/2}^{N/2-1}|{\tilde X}_k|^2|{\tilde Y}_k|^2}
  \label{dcxyerr2}
\end{eqnarray}
and
\begin{eqnarray}
  (\frac{\Delta C_{XY}^\prime}{C_{XY}})^2 & = & (\frac{\Delta C_{XY1}^\prime}{C_{XY}})^2 + (\frac{ <\sigma^2_{XE}>/\sqrt{2N}}{ \sigma^2_X - <\sigma^2_{XE}>})^2 \nonumber \\
  && + (\frac{<\sigma^2_{YE}>/\sqrt{2N}}{\sigma^2_Y- <\sigma^2_{YE}>})^2\nonumber \\
  && 
  \label{dcxyerr3}
\end{eqnarray}
is the estimation of the variation in the presence of measurement
errors.

\begin{figure}[t]
  \begin{center}
    {\includegraphics[width=\linewidth]{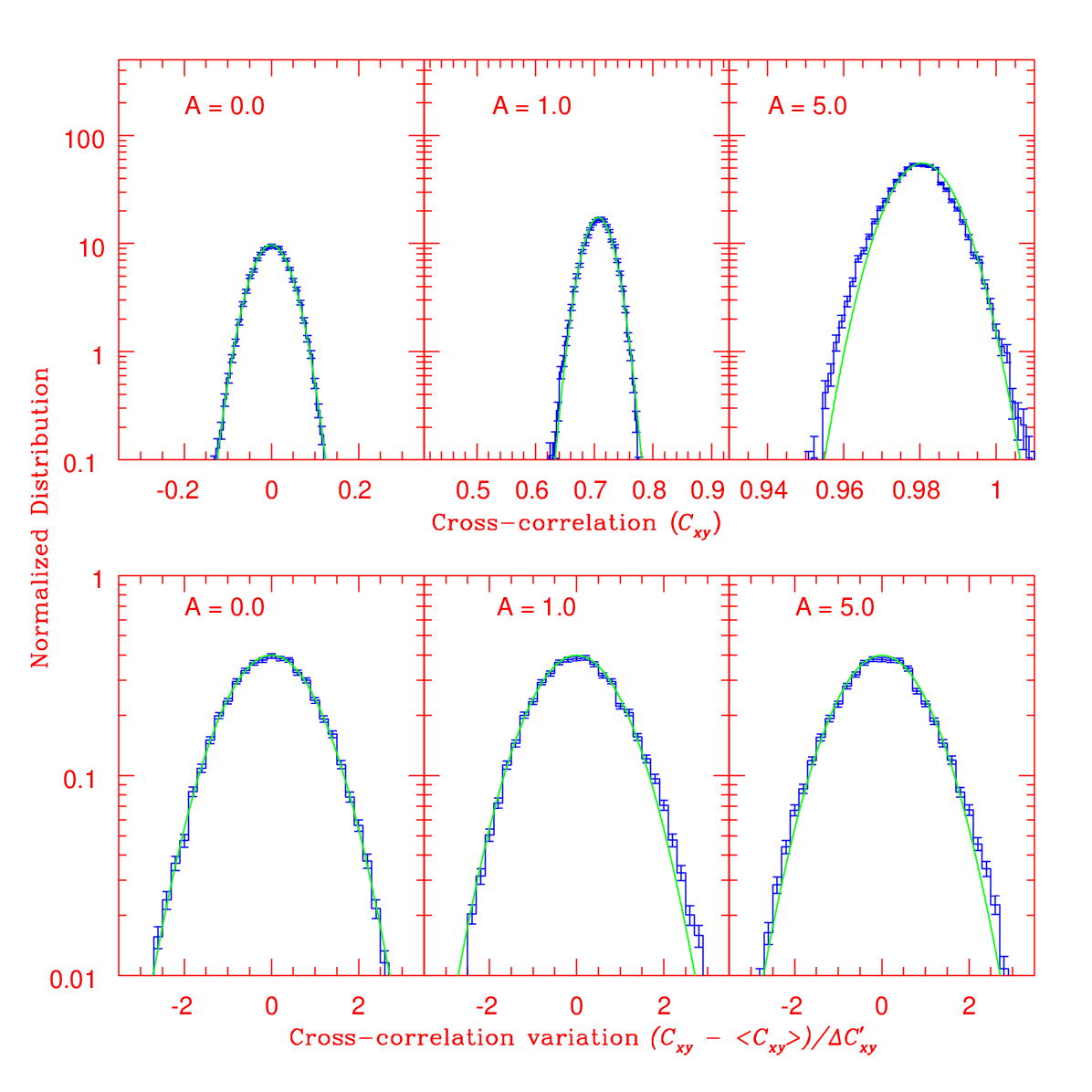}}
  \end{center}
  \caption{Comparison of simulation with analytical results in the
  presence of measurement errors.  19900 pairs of light-curves of
  length $N = 1024$ were created for $X_j = x_j + e_{Xj}$ and $Y_j =
  z_j + Ax_j + e_{Yj}$. $x_j$ and $z_j$ are independent time-series
  generated from a stochastic white noise process (i.e. power spectrum
  index $\alpha = 0$) The measurement errors were simulated from a
  Gaussian distribution such that ${<\sigma^2_{XE}>}/{<\sigma^2_x>} =
  {<\sigma^2_{YE}>}/{<\sigma^2_z>} = 0.5$. Top Panel compares
  normalised histogram $C_{XY}$ with a Gaussian with centroid at the
  expected $<C_{XY}>$ and width given by $\sigma = \Delta C_{xy}$ (Eqn
  \ref{dCxyerr}). Bottom panel shows the histogram of the
  cross-correlation variation $(C_{XY} - <C_{XY}>)/\Delta
  C_{XY}^\prime$. If $\Delta C_{XY}^\prime$ (Eqn \ref{dcxyerr2}),
  which is estimated from the pair of light-curves, is a true measure
  of the variation, then the distribution should be a zero centred
  Gaussian with $\sigma = 1$ (solid line). Note that in the presence
  of measurement errors $<C_{XY}>$ can be greater than one.  }
\end{figure}

\begin{figure}
  \begin{center}
    {\includegraphics[width=\linewidth]{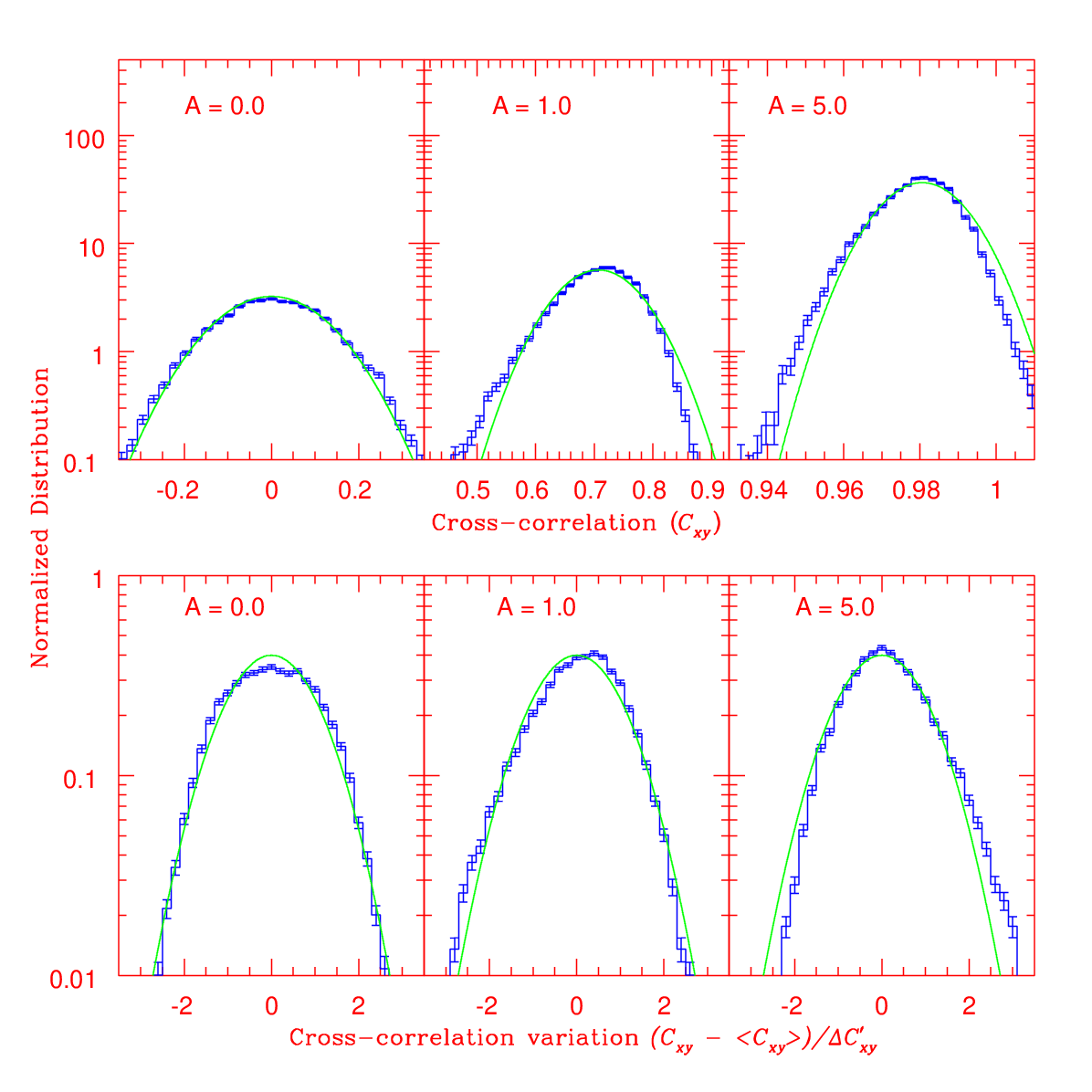}}
  \end{center}
  \caption{ Same as Figure 3, except that $x_j$ and $z_j$ are
  independent time-series generated from a stochastic $1/f$ noise
  process (i.e. power spectrum index $\alpha = 1$).  }
  \label{figa1wnoise}
\end{figure}

To validate the above results we generated 19900 pairs of the
light-curves which obeyed Eqn (\ref{lcerr}). The measurement errors
were generated from a Gaussian distribution with
$${<\sigma^2_{XE}>} / {<\sigma^2_x>} = {<\sigma^2_{YE}>}/{<\sigma^2_z>} =
0.5$$. Figures 3 and 4 show the comparison of the distribution with the
expected Gaussian distribution for power spectral index $\alpha = 0$
and $1$ respectively. As expected the distribution of $C_{XY}$ is
broader in the presence of measurement errors.  Note that in this case
$C_{XY}$ can be greater than unity. The figures illustrate that the
variance estimations reasonably describe the simulated distributions.

The above results are for the case when the measurement errors are
Gaussian distributions.
We have verified that even when the mean
counts per time bin is $\sim 10$, similar results are obtained when
the measurement errors are due to Poisson fluctuations. In order to
correctly propagate the error and obtain Eqn (\ref{dCxyerr}), it is
implicitly assumed that $\Delta \sigma^2_{X,YE} =
(2/\sqrt{N})\sigma^2_{X,YE} << \sigma^2_{X,Y}-\sigma^2_{X,YE}$. Note
that these are also the criteria that any significant variability has
been detected in each of the two light curves. In other words if the
criterion is not satisfied for one of the light curves, this implies
that there is no significantly excess variance than expected from the
measurement errors and hence a cross-correlation analysis cannot be
undertaken.

\section{The cross-correlation function}
\begin{figure}
  \begin{center}
    {\includegraphics[width=\linewidth]{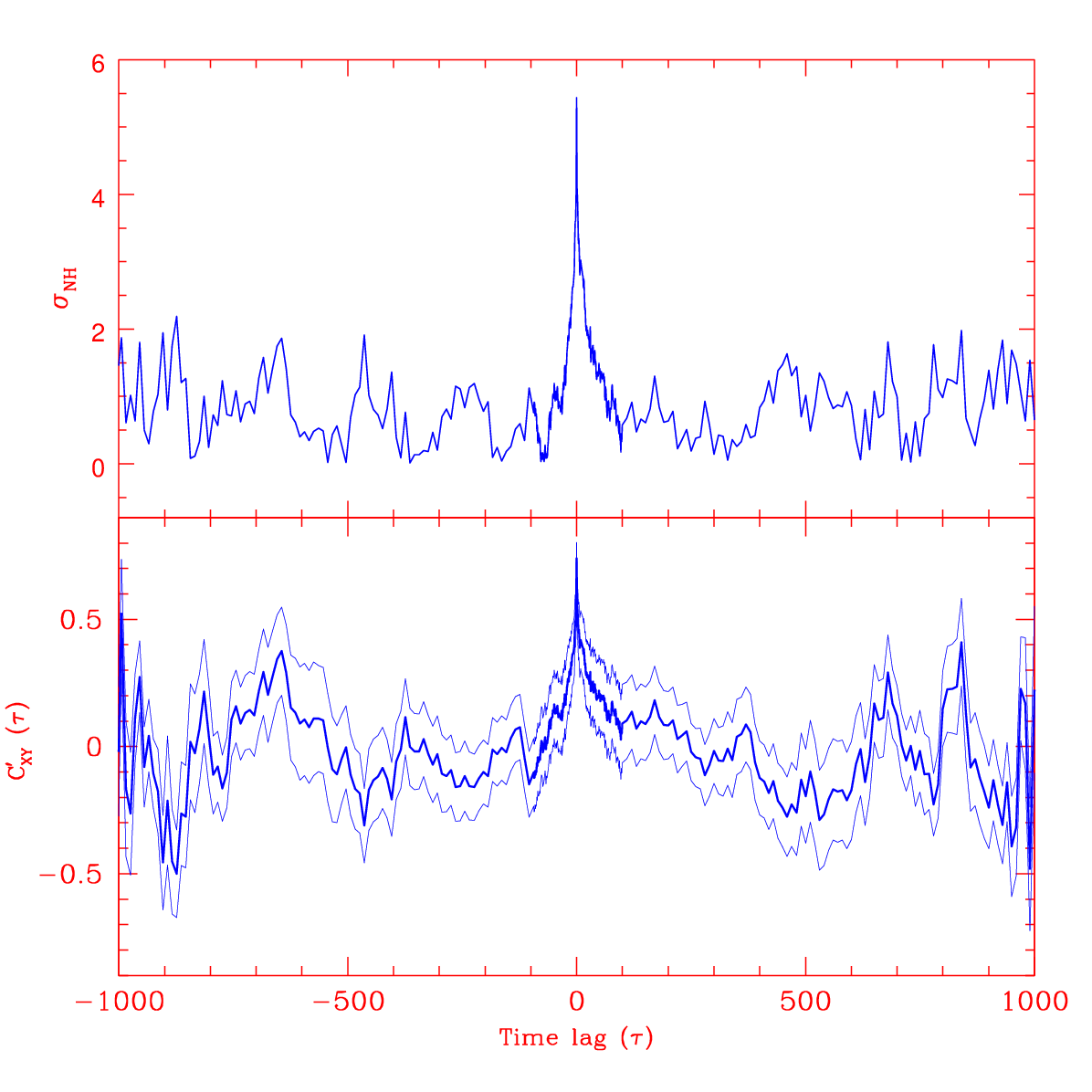}}
  \end{center}
  \caption{Significance and cross-correlation function for two
  simulated light-curves with measurement errors.  Two pairs of
  light-curves of length $N = 1024$ were created for $X_j = x_j +
  e_{Xj}$ and $Y_j = z_j + Ax_j + e_{Yj}$. $x_j$ and $z_j$ are
  independent time-series generated from a stochastic $1/f$ noise
  process (i.e. power spectrum index $\alpha = 1$). The measurement
  errors were simulated from a Gaussian distribution such that
  ${<\sigma^2_{XE}>}/{<\sigma^2_x>} = {<\sigma^2_{YE}>}/{<\sigma^2_z>}
  = 0.5$. The top panel shows the significance $\sigma_{NH} = |C_{XY}|/\Delta C_{XY}^\prime$
  as a function of $\tau$. Note that
  $\sigma_{NH} < 3$ for all $\tau$ except when $|\tau| \sim 0$. The
  bottom panel shows the cross-correlation function, $C_{XY} (\tau)$
  (Thick line), $C_{XY} (\tau) + \Delta C_{XY}^\prime$ and $C_{XY}
  (\tau) - \Delta C_{XY}^\prime$ (Thin lines).  }
\end{figure}

\begin{figure}
  \begin{center}
    {\includegraphics[width=\linewidth]{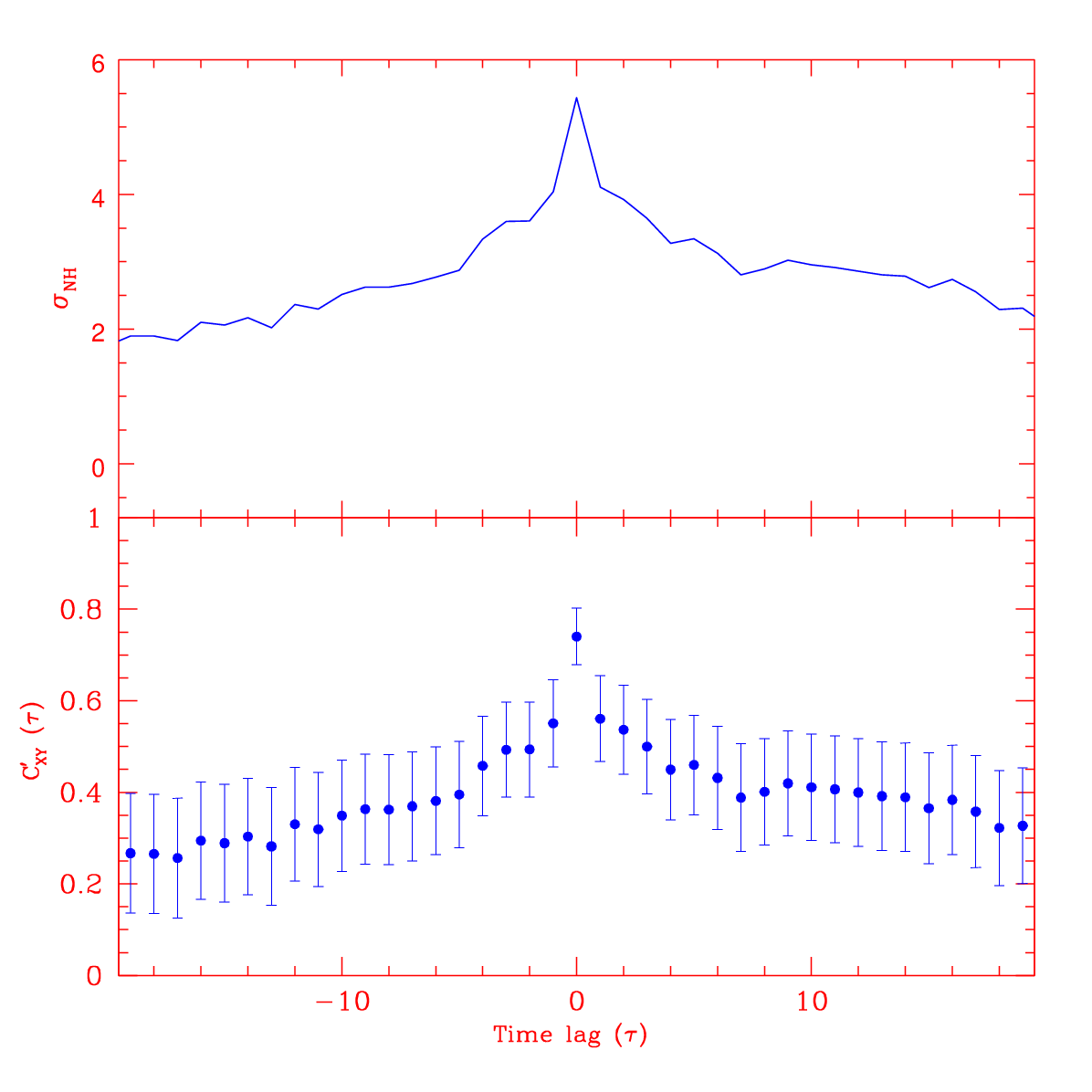}}
  \end{center}
  \caption{ The blown up portion of Fig 5 near $\tau = 0$ shown for
  clarity.  }
\end{figure}

In general two light-curves may be linearly related to each other with
a time lag $\tau$. To investigate such possibilities, one can
calculate the cross-correlation function, $C_{XY} (\tau)$ between a
light curve $X_j$ and the $Y_{j+\tau}$. For light-curves of length
$N$, there will be only $N^\prime (\tau) = N - \tau$ overlapping
terms. $C_{XY} (\tau)$ can be computed based on these $N^\prime
(\tau)$ terms and then the definitions, error analysis of the previous
sections follow through without modifications. Such a definition of
$C_{XY} (\tau)$ has been called {\it locally defined cross-correlation
function} (LDCCF) by \cite{welsh1999reliability} and is different from the standard
one. In the standard definition the length of the original
light-curves $N$ is preserved either by padding the unknown part of
the light with zeros (for the time domain computation) or by repeating
the series (in the Fourier domain computation). Here we consider only
LDCCF for which the analysis mentioned in the earlier section holds.

For every time lag, $\tau$, $\sigma_{NH} = |C_{XY}|/\Delta
C_{XY}^\prime$ needs to be computed to ascertain whether there is any
detectable correlation. In Fig 5, $\sigma_{NH}$ is plotted against
$\tau$ for two light curves generated using $1/f$ stochastic process,
with measurement errors and with $A = 1$. In fact, the two
light-curves are the first pair of light-curves used in the simulation
described in Fig 4. Since $C_{XY}$ is being computed for a large
number of time lags $\tau$, (although they are not independent, see
below), it is prudent to keep a conservative criterion for correlation
detection, $\sigma_{NH} > 3$. It can be seen from the figure that
$\sigma_{NH} < 3$ for all values of $\tau$ except when $\tau \sim 0$.
As seen in the bottom panel of Fig 5, there are couple of peaks in
$C_{XY} (\tau)$ but none of them (except near $\tau = 0$) are
significant.

The $C_{XY} (\tau)$ are not in general independent of each other. This
is clearly seen as an example in Fig 6 where the region near $\tau =
0$ has been expanded for clarity. For $|\tau| < 3$, $\sigma_{NH} > 3$
and the dispersion of $C_{XY} (\tau)$ is much less than what is expected from
the error bars $\Delta C_{XY}^\prime$. This shows that it is unpreferable
to rebin $C_{XY} (\tau)$ in $\tau$ space. Instead if there is a
justification to rebin in time, then the original light-curves should
be rebinned and not $C_{XY} (\tau)$. The data clearly shows a peak of
$C_{XY} (\tau)$ at $\tau = 0$ indicating a time-lag consistent with
zero. However, it is difficult to justify any error measurement on
this time lag. In practice, one can represent the peak of the function
as a Gaussian and take its width as an representative error for the
lag. However, there are several un-attractive features of this
technique. First, the width of $C_{XY} (\tau)$ represents the
auto-correlation of the coherent signal rather than any error on the
measured time-lag. Second, since the errors on $C_{XY} (\tau)$ are
correlated a formal fit is not allowed. Finally, the Gaussian fit will
in general depend on the number of points used to represent the
``peak'' of $C_{XY} (\tau)$. A justifiable technique to compute the
error on the time-lag is required.

\section{The cross-Correlation Phasor}

\begin{figure}
  \begin{center}
    {\includegraphics[width=\linewidth]{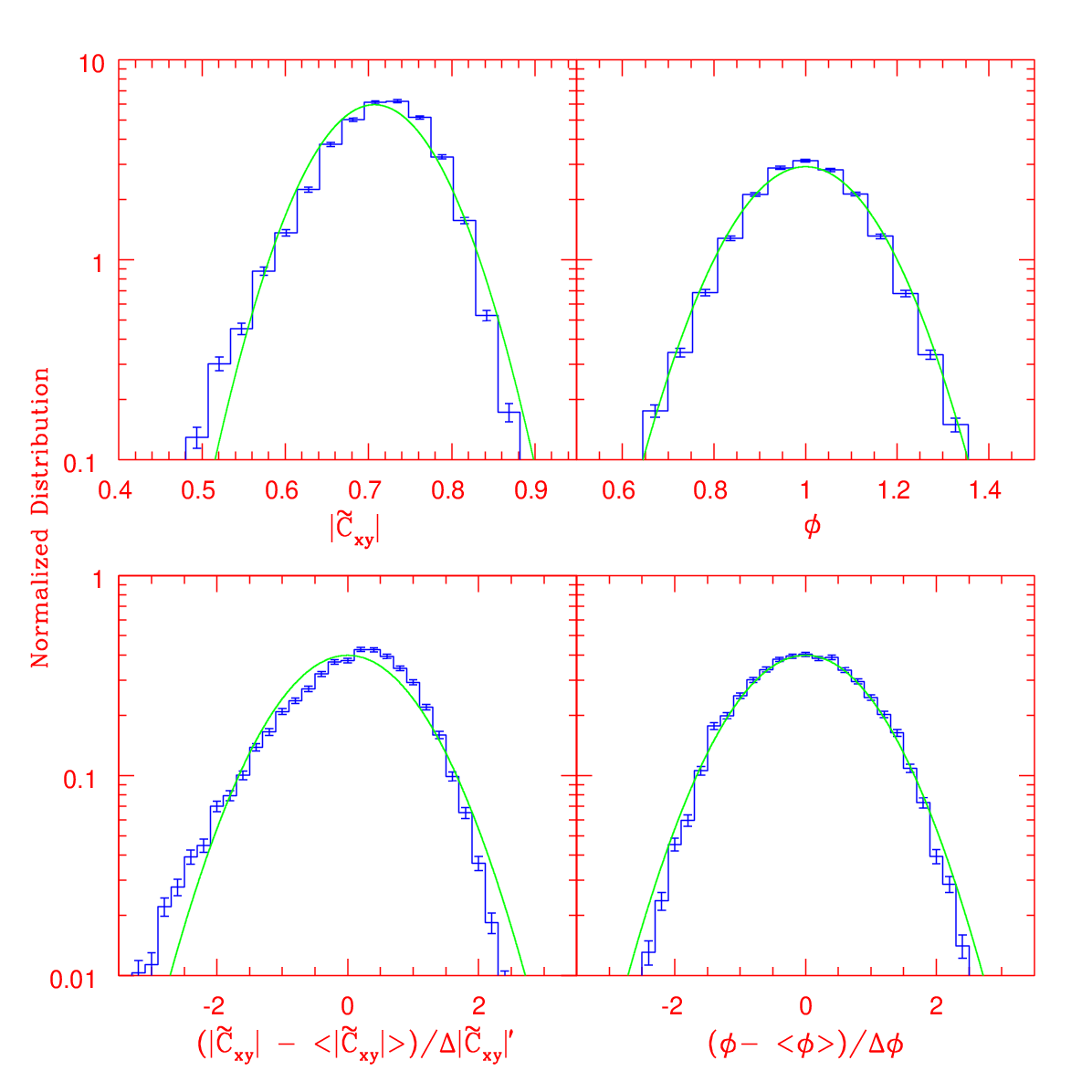}}
  \end{center}
  \caption{Comparison of simulation with analytical results. 19900
  pairs of light-curves of length $N = 1024$ were created for $X_j =
  x_j$ and $Y_j = z_j + Ae^{i<\phi>}x_j$. $x_j$ and $z_j$ are
  independent time-series generated from a stochastic $1/f$ noise
  process (i.e. power spectrum index $\alpha = 1$). The values of the
  parameters used for the simulations are $A = 1$ and $<\phi> =
  1$. Solid lines are the expected distribution for the error
  estimates $\Delta |\tilde C_{XY}|$ and $\Delta \phi$.  $\Delta
  |\tilde C_{XY}^\prime|$ is given by Eqn (\ref{dcxyerr3}) except that
  $C_{XY}$ is replaced by $|\tilde C_{XY}|$ while $\Delta \phi$ is
  given by Eqn (\ref{dphi})}.
\end{figure}

The cross-correlation phasor can be defined as
\begin{equation}
  \tilde C_{XY}  = \frac{\tilde c_{XY}}{\sqrt{\sigma^2_{X} \sigma^2_Y}}
\end{equation}
where
\begin{equation}
  \tilde c_{XY}   = \frac{2}{N^2}\sum_{k=1}^{N/2-1} {\tilde X}_k {\tilde Y}_k^*
  \label{smallcxypha}
\end{equation}
Here the summation of phasor differs from that of the cross-correlation by the fact that
for the phasor the summation is only over positive frequencies.
They are related as $C_{XY} = Re (\tilde C_{XY})$.

For partially correlated light curves with no phase lag, the ensemble
average $< Im(\tilde C_{XY}) > = 0$ and the cross-correlation is given
by $<Re (\tilde C_{XY})> = <C_{XY}>$. By definition, the deviation of
$Re(\tilde c_{XY})$ is the same as for $c_{XY}$ whereas the deviation 
of $Im(\tilde c_{XY})$, is only due to the incoherent
parts of the light curves and is given by the equation
\begin{equation}
  \Delta Im(\tilde c_{XY})  = \Delta c_{XY} \sqrt{1 - |\tilde C_{XY}|^2}
\end{equation}

If there is an intrinsic phase difference, $<\phi>$ between the two
light curves, then the ensemble average of $\tilde C_{XY}$ will be a
complex quantity given by $<|\tilde C_{XY}|>e^{i<\phi>}$.  The standard
deviation of $|\tilde C_{XY}|$ from $< | \tilde C_{XY}| >$ can
be estimated by Eqn (\ref{dcxyerr3}) except that $C_{XY}$ is to be
replaced by $|\tilde C_{XY}|$. The phase difference between the two
light curves can be estimated as
\begin{equation}
  \hbox {sin} \phi = \frac{Im(\tilde c_{XY})}{|\tilde c_{XY}|}
\end{equation}
whose error, for small values of $\phi$, can be estimated to be
\begin{equation}
  \Delta \phi = \frac{\Delta c_{XY}}{|\tilde c_{XY}|} \sqrt{1 - |\tilde C_{XY}|^2}
  \label{dphi}
\end{equation}

To validate the above results we simulated the same set of light
curves used for Figure (\ref{figa1wnoise}) i.e. using 19900 pairs of
light curves with measurement errors and generated from a stochastic
$1/f$ noise process. We introduced a phase difference of $\phi = 1.0$
between the coherent parts of the light curves. The histograms of
$|\tilde C_{XY}|$ and $\phi$ (and their deviations) are plotted
against the expected estimates in Figure 7.

If the coherent parts of the light curves have a time-lag, $\tau$,
between them, then the cross-correlation phasor will have a non-zero
phase. One can constrain the time-lag by shifting one of the light
curves in time till the cross-correlation phase, $\phi = 0$.  In other
words, by defining a cross-correlation phasor function,
\begin{equation}
  \tilde C_{XY} (\tau)   = \frac{2}{N^2 \sqrt{\sigma^2_{X} \sigma^2_Y}}\sum_{k=1}^{N/2-1} {\tilde X}_k {\tilde Y}_k^* e^{{ik\tau}/{N}}
  \label{CXYfunc}
\end{equation}
one can obtain $\tau$ such the phase of $\tilde C_{XY} (\tau)$,
$\phi(\tau) = 0$. The error on $\tau$ can be estimated by considering
the range of $\tau$ for which $\phi(\tau^\prime)\pm \Delta \phi $ is
consistent with zero.  Note that $\tau$ need not be an integer and
hence time-lags less than the time resolution of the light curves can
be ascertained for good quality data.

The above analysis is valid only when there is a detected correlation
between the two light curves. To ascertain whether there is a
correlation (with phase lag) between the two, one needs to consider
both the real and imaginary parts of $\tilde c_{XY}$ and compare with
$\Delta c_{XY}$. While one can compute the joint probabilities, a more
prudent and simpler approach is to demand that a correlation is
detected only if $|\tilde C_{XY}|/\Delta C_{XY} > 3$. If the condition
is not satisfied one can put an upper limit on the correlation as $3
\Delta c_{XY}/ \sqrt{\sigma^2_{X} \sigma^2_Y}$.

\section{Algorithm to compute cross-correlation, phase and time lags}

Based on the results of the earlier sections, we present here a
complete self-contained description of the algorithm to compute and
estimate errors for cross-correlation, phase and time lag between two
light curves. It is assumed that there are two time series $X_j$ and
$Y_j$ of length $N$ and for each data there are associated known
measurement errors $e_{Xj}$ and $e_{Yj}$.  \smallskip

\noindent {\it Step 1: Calculate variances for the light curves.}
Compute the Fourier transforms of each light curve using
\begin{equation}
  \tilde{X}_{k} = \sum_{j=0}^{N-1} {X}_{j} \exp{(2\pi ijk/N)}
\end{equation}
and similarly for $\tilde{Y}_{k}$. Compute intrinsic variances,
\begin{equation}
  \sigma^2_{XI}  = \frac{2}{N^2}\sum_{k=1}^{N/2-1} |{\tilde X}_k|^2 - \sigma^2_{XE}
\end{equation}
where $\sigma^2_{XE} = \frac{1}{N} \sum_{j=0}^{N-1} e_{Xj}^2$ and
similarly for $\sigma^2_{YI}$.  If either $\sigma^2_{XI}< 2\Delta
\sigma^2_{XE} = (4/\sqrt{N})\sigma^2_{XE}$ or $\sigma^2_{YI}<
(4/\sqrt{N})\sigma^2_{YE}$, then no significant variation has been
detected in one of the light curves and further analysis is not
possible. In such cases, an upper limit of $(4/\sqrt{N})\sigma^2_{XE}$
can be put for any intrinsic variation. If there is significant
variation detected, then the total error (both stochastic and
measurement) on the variance is
\begin{equation}
  \Delta \sigma^2_{XI} = \sqrt {\frac{2}{N^4}\sum_{k=1}^{N/2-1} (|{\tilde X}_k|^2)^2 }
\end{equation}
of which the measurement uncertainty contributes
\begin{equation}
  \Delta \sigma^2_{XM} = \sqrt {(\Delta \sigma^2_{XT})^2 - \frac{2}{N^4}\sum_{k=1}^{N/2-1} (|{\tilde X}_k|^2 -|{\tilde X}_M|^2  )^2 }
\end{equation}
where $|{\tilde X}_M|^2 = N \sigma^2_{XE}$ and is independent of $k$.
\bigskip

\noindent {\it Step 2: Compute the cross-Correlation.} The
non-normalised cross-correlation phasor is
\begin{equation}
  \tilde c_{XY}   = \frac{2}{N^2}\sum_{k=1}^{N/2-1} {\tilde X}_k {\tilde Y}_k^*
\end{equation}
and its error is given by
\begin{equation}
  (\Delta c_{XY})^2 = \frac{1}{N^4}\sum_{k=-N/2}^{N/2-1} |{\tilde X}_k|^2|{\tilde Y}_k|^2
\end{equation}
Check if $|\tilde c_{XY}|/\Delta c_{XY} > 3$. If not then no
correlation is detected between the two light curves and the upper
limit on the cross-correlation is $3 \Delta c_{xy}/
\sqrt{\sigma^2_{XI} \sigma^2_{YI}}$.  If the condition is satisfied
(i.e. the correlation is detected) then the cross-correlation is
\begin{equation}
  |\tilde C_{XY}|  = \frac{|\tilde c_{XY}|}{\sqrt{\sigma^2_{XI} \sigma^2_{YI}}}
\end{equation}
with error
\begin{equation}
  (\frac{\Delta |\tilde C_{XY}|}{|\tilde C_{XY}|})^2  =  (\frac{\Delta |\tilde C_{XY1}|}{|\tilde C_{XY}|})^2 + (\frac{ \sigma^2_{XE}}{ \sqrt{2N}\sigma^2_{XI}})^2 + (\frac{\sigma^2_{YE}}{\sqrt{2N}\sigma^2_{YI}})^2 
\end{equation}
where
\begin{equation}
  \Delta |\tilde C_{XY1}|  =  \frac{(1 - C_{XY}^2) \Delta c_{XY}}{ \sqrt{\sigma^2_{XI} \sigma^2_{YI}}} 
\end{equation}
\smallskip

\noindent {\it Step 3: Compute the phase.} The phase is given by
\begin{equation}
  \hbox {sin} \phi = \frac{Im(\tilde c_{XY})}{|\tilde c_{XY}|}
\end{equation}
whose error can be estimated to be
\begin{equation}
  \Delta \phi = \frac{\Delta c_{XY}}{|\tilde c_{XY}|} \sqrt{1 - |\tilde C_{XY}|^2}
\end{equation}
\smallskip

\noindent {\it Step 4: Compute the time delay between the light
curves}. Define
\begin{equation}
  \tilde C_{XY} (\tau^\prime)   = \frac{2}{N^2 \sqrt{\sigma^2_{XI} \sigma^2_{YI}}}\sum_{k=1}^{N/2-1} {\tilde X}_k {\tilde Y}_k^* e^{{ik\tau^\prime}/{N}}
\end{equation}
and solve for $\phi (\tau) = 0$ to get an estimate of the time delay
$\tau$.  The error on $\tau$, $\Delta \tau$ is to be estimated by
considering the range of $\tau^\prime$ for which $\phi(\tau^\prime)\pm
\Delta \phi (\tau^\prime)$ is consistent with zero. Compute the
significance of the cross-correlation $|\tilde c_{XY}|/\Delta c_{XY}$
at the two limits $\tau^\prime = \tau \pm \Delta \tau$ and consider
the limits to be bona-fide if the significance is $ > 3$, otherwise
report that the particular limit on $\tau$ cannot be obtained.

\smallskip
\noindent {\it Step 5: For multiple light curves or for a lightcurve
divided in to segments} find the weighted average of $\sigma^2_{XI}$,
$\sigma^2_{YI}$ and $\tilde c_{XY}$, using their error estimates as
weights. Then if the cross-correlation is significant, find the phase
and time lags as in stjpg 3 and 4 above.

\section{Application to AGN light curves}

\begin{figure}
  \begin{center}
    {\includegraphics[width=\linewidth]{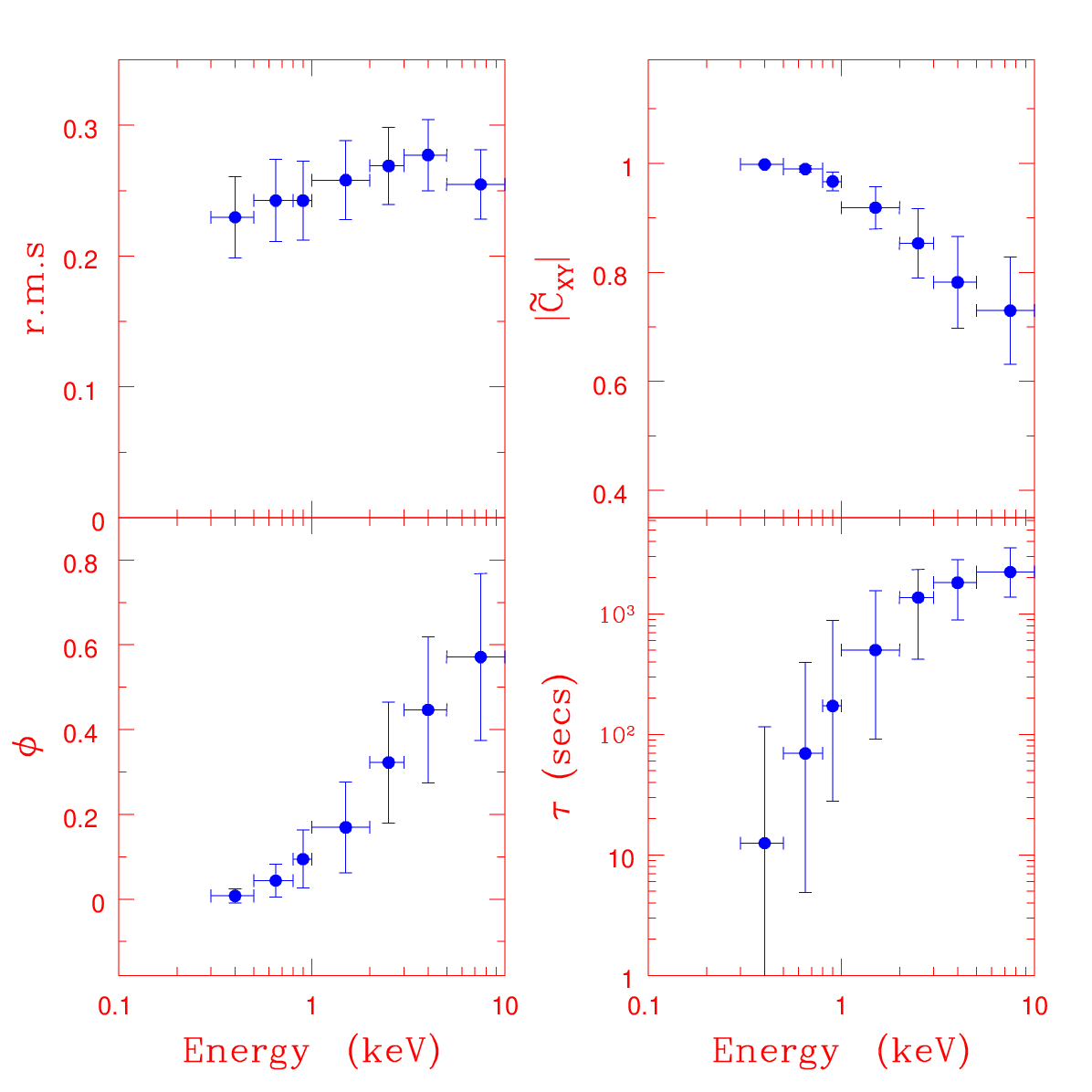}}
  \end{center}
  \caption{ The variability property of Akn~564.  Lightcurves of the
  source in different energy bands were used for the analysis. The
  time-bin for the light curves is 64 seconds and the number of data
  points is 1426. The rms, cross-correlation ($|\tilde C_{XY}|$), the
  phase difference ($\phi$) and the time lag ($\tau$) are plotted with
  energy. The reference energy band is $0.2$-$0.3$ keV.  }
  \label{figAKN1}
\end{figure}

\begin{figure}
  \begin{center}
    {\includegraphics[width=\linewidth]{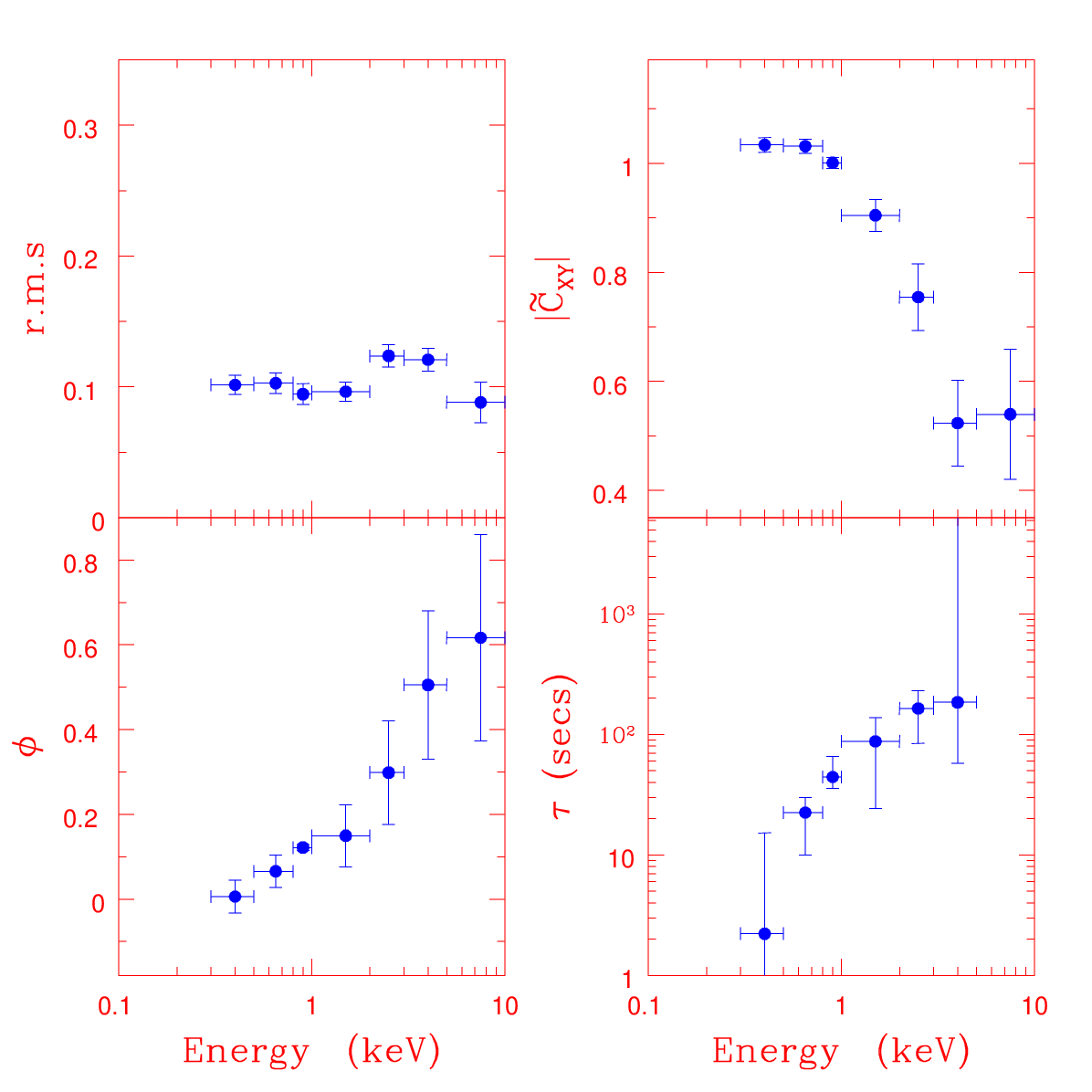}}
  \end{center}
  \caption{ Same as Figure \ref{figAKN1}, except that the light curves
  were divided into ten segments and the results averaged.  }
  \label{figAKN2}
\end{figure}

\begin{figure}
  \begin{center}
    {\includegraphics[width=\linewidth]{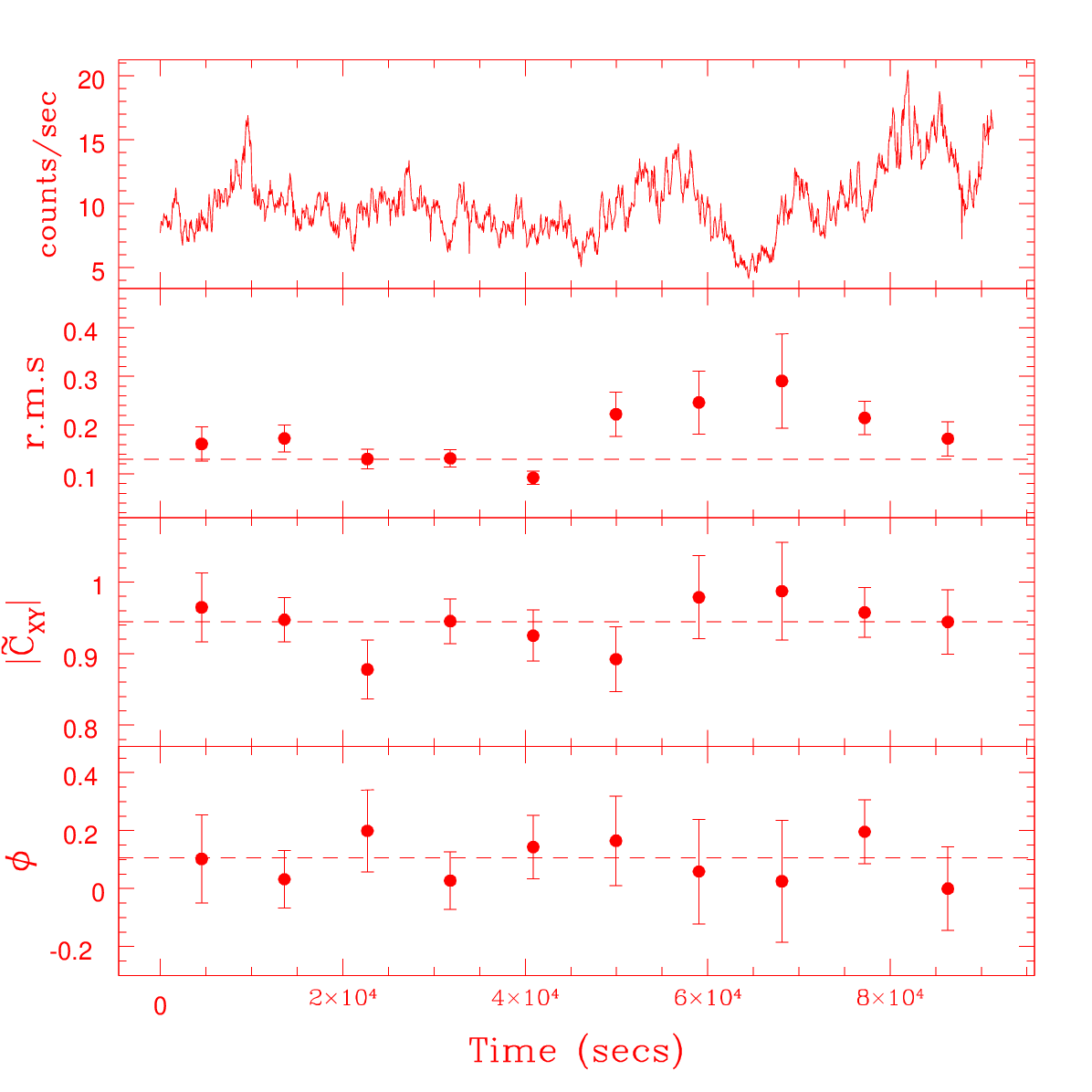}}
  \end{center}
  \caption{ Checking the stationarity of the X-ray lightcurve of
  Akn~564. The complete lightcurve in the $1$-$2$ keV band (shown in
  the top panel) has been divide into ten segments. For each segment
  the r.m.s is shown with errors in the second panel. The dashed line
  represents the average value. The cross-correlations, $|\tilde
  C_{XY}|$), and phase lags, $\phi$ between $0.2-1.$ and $1$-$2$ keV
  bands for each segment are shown in the bottom two panels. The
  dashed lines represent the average values.  It can seen that the
  cross-correlation and the phase lag are consistent with being a
  constant showing that the system is stationary in these time-scales.
  }
  \label{figAKN3}
\end{figure}

To test and validate the effectiveness of the scheme, we analyse the
lightcurve of a well studied Active Galactic Nucleus, Akn~564.  The
temporal and spectral properties of the source was studied using an
{\it XMM-Newton} observation of the source by \cite{dewangan2007investigation}.  They
computed the cross-correlation function for different energy bands and
estimated a possible time-lag between the hard and soft bands to be
$\sim 1768$ secs using the peak of the function as a measure. Using
Monte Carlo simulations they estimated an error on the time lag to be
$\sim 100$ secs due to measurement error. \cite{arevalo2006spectral} and
\cite{mchardy2007discovery} computed time-lags as function of frequency for {\it
ASCA} and {\it XMM-Newton} observations of the source and found that
there is a sharp drop in time lag for frequencies greater than
$10^{-4}$ Hz.

We extracted light curves of the source using the {\it XMM-Newton}
observation, in different energy bands.  Details of the extraction
process are given in \cite{dewangan2007investigation}. The usable continuous time duration
for the observation is for $\sim 10^5$ secs.  Our motivation here is
not to analyse in detail the temporal properties of the source and
their physical interpretation, but instead to show as an example and
validate the method described in this work. Thus, while finer time
binning of the data is possible, we restrict our analysis to $64$ sec
bins, which resulted in light curves with length $N = 1426$.  Figure
\ref{figAKN1} shows the results of our analysis. Note that the
cross-correlation, phase and time-lag are well constrained as a
function of energy. Figure \ref{figAKN2} shows the results of the
analysis when the light curves were divided into ten segments and the
results averaged as described in the last section. Note that again the
physical quantities are well constrained and while the phase
difference is relatively unchanged between the two analysis, the
time-lag decreases by nearly an order of magnitude. This is consistent
with earlier results that the time-lag decreases with increasing
Fourier frequency \cite{papadakis2001frequency, vaughan2003x}.

Splitting the lightcurve into segments and taking the average assumes
that the during the time-scale, the source was stationary. This can be
now explicitly tested using the analytical error estimates for each
segment. This is demonstrated in Fig \ref{figAKN3}, where for each ten
segments, the r.m.s, cross-correlations and phase-lags (between the
energy bands $0.2$-$1$ and $1$-$2$ keV) are shown. The
cross-correlations and phase lags are consistent with being a constant
equal to the averaged value (shown as a dashed line). Formally the
$\chi^2/dof$ for the data points to be constant are $5.3/9$ and
$3.3/9$ for the cross-correlation and phase lag respectively. The
slightly lower value of $\chi^2$ than expected indicates only a slight
overestimation of the error bars, especially for the phase lags. This
is probably because for each segment, the error on the phase lag
$\Delta \phi$ is large and hence the error distribution maybe slightly
different than a Gaussian.  Nevertheless, the figure clearly shows
that not only can stationarity be tested but also reconfirms that the
error estimates are reliable. Perhaps, not surprisingly, given the
shape of the total light curve, the r.m.s of the $1$-$2$ keV energy
band is formally not consistent with being a constant, with a
$\chi^2/dof = 28/10$. This is primarily due to the fifth segment where
the r.m.s and its error is small. Note that the error is computed by
assuming that the power spectrum of the segment is representative of
the ensemble average. Perhaps a more prudent approach would be to
estimate the error on the r.m.s using the averaged power spectrum
rather than for each individual segment.  However, whether such
deviations are a significant indication of departure from stationarity
is arguable and subjective.  Hence, we recommend that only large
deviations (for e.g. $\chi2/dof > 5$) should be taken as serious
evidence for non-stationarity.

\section{Summary and discussion}
\label{sect:discussion}
In this work an estimate for the significance and error on the
cross-correlation, phase and time lag between two light curves is
presented. The error estimates take into account the stochastic
fluctuations of the lightcurve as well as any known measurement
errors. The technique has been verified using simulations of light
curves generated from both white and $1/f$ stochastic processes with
and without intrinsic correlation between them.  The entire analysis
consists of five algorithmic stjpg which are described in \textsection 6. The
technique is ideally suited for short light curves of length $N \sim
1000$ and is an improvement over earlier methods which were based on
numerically expensive simulations or by dividing the data into number
of segments to find the variance.

The estimate presented in this work is based on several assumptions and
hence is reliable only when they are valid. We emphasize this point
by enumerating some of the main assumptions.

\noindent $\bullet$ Both the light curves have been generated from
stochastic processes. Technically, this means that the phase of the
different Fourier components are unrelated to each other i.e. $ <
\tilde X_k \tilde X_l > = \delta_{lk}$.  This assumption will be
violated if the generation mechanism is a non-linear one.  In general,
it is difficult to ascertain the degree of non-linearity in a short
lightcurve and it requires sensitive analysis like Bi-coherence
measure and/or non-linear time series analysis. Thus, in most cases,
the stochastic nature of the light curves have to be assumed.  It is
prudent to be aware that this assumption has been made and its
validity is unknown, like for example, for the prompt emission of
Gamma-ray bursts. A simple case where this assumption will be violated
is if the power spectra have dominant harmonic features, where the
power in the harmonics is comparable to that of the primary.

\noindent $\bullet$ The measurement errors are uncorrelated and have
Gaussian distributions. The essential assumption is that the power
spectrum of the measurement errors is independent of frequency (i.e. a
white noise) and their phases are independent of each other. If the
power spectrum has a different shape, then the appropriate changes
have to be made and the basic results of this work need to be
re-derived. For most practical purposes if the measurement errors are
known, they usually are Gaussian distributions and hence this
assumption is valid. If there are unknown systematic errors in the
light curves then of course the analysis will not be
applicable. Poisson distributions have the white noise property but in
general the phases of the different Fourier components may be
related. We have verified that for counts per time bin $\sim 10$, the
results of this analysis is valid. For counts less than that, caution
is advised. However, for such low counts, meaningful results can only
be obtained for long time series and it may better to obtain frequency
dependent coherence and lag measurements.

\noindent $\bullet$ The light curves are evenly sampled without
gaps. For unevenly sampled data the cross-correlation can be estimated
\cite{edelson1988discrete}, but there does not seem to be an analytical way to
estimate the significance and error. One needs to use either Monte
Carlo simulations or more practically estimate the error by dividing
the light curves into several segments and finding their variance.

\noindent $\bullet$ The light curves are stationary. As shown in the
example of the light curves of Akn~564, this assumption can be tested
by dividing the light curves into segments and checking whether the
r.m.s, cross-correlation and phase lags are consistent to be a
constant for different segments.

While this technique is useful for short duration light curves,
coherence and frequency dependent time lags provide naturally more
information and should be preferentially computed for long data
streams. This technique may not be unique or optimal and hence there
is a possibility and need for development of better methods provided
they give robust and physically interpretable results. Finally, while
cross-correlation, phase and time lags provide a quantitative measure
of the system, their physical interpretation has to be done in terms
of the physical geometrical and radiative model assumed for the
system. 


\section{Acknowledgement}
This work has made use of observational data obtained with XMM-Newton,  an  ESA  science  mission  with  instruments  and  
contributions  directly  funded  by ESA Member States and the USA (NASA). AB would like to thank the Inter-University Centre 
for Astronomy and Astrophysics, Pune for associateship programme. 


\begin{thebibliography}{16}
\expandafter\ifx\csname natexlab\endcsname\relax\def\natexlab#1{#1}\fi
\providecommand{\url}[1]{\texttt{#1}}
\providecommand{\href}[2]{#2}
\providecommand{\path}[1]{#1}
\providecommand{\DOIprefix}{doi:}
\providecommand{\ArXivprefix}{arXiv:}
\providecommand{\URLprefix}{URL: }
\providecommand{\Pubmedprefix}{pmid:}
\providecommand{\doi}[1]{\href{http://dx.doi.org/#1}{\path{#1}}}
\providecommand{\Pubmed}[1]{\href{pmid:#1}{\path{#1}}}
\providecommand{\bibinfo}[2]{#2}
\ifx\xfnm\relax \def\xfnm[#1]{\unskip,\space#1}\fi
\bibitem[{Ar{\'e}valo et~al.(2006)Ar{\'e}valo, Papadakis, Uttley, McHardy and
  Brinkmann}]{arevalo2006spectral}
\bibinfo{author}{Ar{\'e}valo, P.}, \bibinfo{author}{Papadakis, I.},
  \bibinfo{author}{Uttley, P.}, \bibinfo{author}{McHardy, I.},
  \bibinfo{author}{Brinkmann, W.}, \bibinfo{year}{2006}.
\newblock \bibinfo{title}{Spectral-timing evidence for a very high state in the
  narrow-line seyfert 1 ark 564}.
\newblock \bibinfo{journal}{Monthly Notices of the Royal Astronomical Society}
  \bibinfo{volume}{372}, \bibinfo{pages}{401--409}.
\bibitem[{Bartlett(1955)}]{bartlett1955introduction}
\bibinfo{author}{Bartlett, M.S.}, \bibinfo{year}{1955}.
\newblock \bibinfo{title}{An Introduction to stochastic processes: with special
  reference to methods and applications}.
\newblock \bibinfo{publisher}{The University Press}.
\bibitem[{Chatfield(2016)}]{chatfield2016analysis}
\bibinfo{author}{Chatfield, C.}, \bibinfo{year}{2016}.
\newblock \bibinfo{title}{The analysis of time series: an introduction}.
\newblock \bibinfo{publisher}{CRC press}.
\bibitem[{Dewangan et~al.(2007)Dewangan, Griffiths, Dasgupta and
  Rao}]{dewangan2007investigation}
\bibinfo{author}{Dewangan, G.}, \bibinfo{author}{Griffiths, R.},
  \bibinfo{author}{Dasgupta, S.}, \bibinfo{author}{Rao, A.},
  \bibinfo{year}{2007}.
\newblock \bibinfo{title}{An investigation of the origin of soft x-ray excess
  emission from ark 564 and mrk 1044}.
\newblock \bibinfo{journal}{The Astrophysical Journal} \bibinfo{volume}{671},
  \bibinfo{pages}{1284}.
\bibitem[{Edelson and Krolik(1988)}]{edelson1988discrete}
\bibinfo{author}{Edelson, R.}, \bibinfo{author}{Krolik, J.},
  \bibinfo{year}{1988}.
\newblock \bibinfo{title}{The discrete correlation function-a new method for
  analyzing unevenly sampled variability data}.
\newblock \bibinfo{journal}{The Astrophysical Journal} \bibinfo{volume}{333},
  \bibinfo{pages}{646--659}.
\bibitem[{Keshner(1982)}]{keshner19821}
\bibinfo{author}{Keshner, M.S.}, \bibinfo{year}{1982}.
\newblock \bibinfo{title}{1/f noise}.
\newblock \bibinfo{journal}{Proceedings of the IEEE} \bibinfo{volume}{70},
  \bibinfo{pages}{212--218}.
\bibitem[{Van~der Klis(1989)}]{van1989fourier}
\bibinfo{author}{Van~der Klis, M.}, \bibinfo{year}{1989}.
\newblock \bibinfo{title}{Fourier techniques in x-ray timing}, in:
  \bibinfo{booktitle}{Timing Neutron Stars}. \bibinfo{publisher}{Springer}, pp.
  \bibinfo{pages}{27--69}.
\bibitem[{Leahy et~al.(1983)Leahy, Darbro, Elsner, Weisskopf, Kahn, Sutherland
  and Grindlay}]{leahy1983searches}
\bibinfo{author}{Leahy, D.}, \bibinfo{author}{Darbro, W.},
  \bibinfo{author}{Elsner, R.}, \bibinfo{author}{Weisskopf, M.},
  \bibinfo{author}{Kahn, S.}, \bibinfo{author}{Sutherland, P.},
  \bibinfo{author}{Grindlay, J.}, \bibinfo{year}{1983}.
\newblock \bibinfo{title}{On searches for pulsed emission with application to
  four globular cluster x-ray sources-ngc 1851, 6441, 6624, and 6712}.
\newblock \bibinfo{journal}{The Astrophysical Journal} \bibinfo{volume}{266},
  \bibinfo{pages}{160--170}.
\bibitem[{McHardy et~al.(2007)McHardy, Ar{\'e}valo, Uttley, Papadakis, Summons,
  Brinkmann and Page}]{mchardy2007discovery}
\bibinfo{author}{McHardy, I.}, \bibinfo{author}{Ar{\'e}valo, P.},
  \bibinfo{author}{Uttley, P.}, \bibinfo{author}{Papadakis, I.},
  \bibinfo{author}{Summons, D.}, \bibinfo{author}{Brinkmann, W.},
  \bibinfo{author}{Page, M.}, \bibinfo{year}{2007}.
\newblock \bibinfo{title}{Discovery of multiple lorentzian components in the
  x-ray timing properties of the narrow line seyfert 1 ark 564}.
\newblock \bibinfo{journal}{Monthly Notices of the Royal Astronomical Society}
  \bibinfo{volume}{382}, \bibinfo{pages}{985--994}.
\bibitem[{Nowak et~al.(1999)Nowak, Vaughan, Wilms, Dove and
  Begelman}]{nowak1999rossi}
\bibinfo{author}{Nowak, M.A.}, \bibinfo{author}{Vaughan, B.A.},
  \bibinfo{author}{Wilms, J.}, \bibinfo{author}{Dove, J.B.},
  \bibinfo{author}{Begelman, M.C.}, \bibinfo{year}{1999}.
\newblock \bibinfo{title}{Rossi x-ray timing explorer observation of cygnus
  x-1. ii. timing analysis}.
\newblock \bibinfo{journal}{The Astrophysical Journal} \bibinfo{volume}{510},
  \bibinfo{pages}{874}.
\bibitem[{Papadakis et~al.(2001)Papadakis, Nandra and
  Kazanas}]{papadakis2001frequency}
\bibinfo{author}{Papadakis, I.}, \bibinfo{author}{Nandra, K.},
  \bibinfo{author}{Kazanas, D.}, \bibinfo{year}{2001}.
\newblock \bibinfo{title}{Frequency-dependent time lags in the x-ray emission
  of the seyfert galaxy ngc 7469}.
\newblock \bibinfo{journal}{The Astrophysical Journal Letters}
  \bibinfo{volume}{554}, \bibinfo{pages}{L133}.
\bibitem[{Peterson et~al.(1998)Peterson, Wanders, Horne, Collier, Alexander,
  Kaspi and Maoz}]{peterson1998uncertainties}
\bibinfo{author}{Peterson, B.M.}, \bibinfo{author}{Wanders, I.},
  \bibinfo{author}{Horne, K.}, \bibinfo{author}{Collier, S.},
  \bibinfo{author}{Alexander, T.}, \bibinfo{author}{Kaspi, S.},
  \bibinfo{author}{Maoz, D.}, \bibinfo{year}{1998}.
\newblock \bibinfo{title}{On uncertainties in cross-correlation lags and the
  reality of wavelength-dependent continuum lags in active galactic nuclei}.
\newblock \bibinfo{journal}{Publications of the Astronomical Society of the
  Pacific} \bibinfo{volume}{110}, \bibinfo{pages}{660}.
\bibitem[{Timmer and Koenig(1995)}]{timmer1995generating}
\bibinfo{author}{Timmer, J.}, \bibinfo{author}{Koenig, M.},
  \bibinfo{year}{1995}.
\newblock \bibinfo{title}{On generating power law noise.}
\newblock \bibinfo{journal}{Astronomy and Astrophysics} \bibinfo{volume}{300},
  \bibinfo{pages}{707}.
\bibitem[{Vaughan et~al.(2003a)Vaughan, Edelson, Warwick and
  Uttley}]{vaughan2003characterizing}
\bibinfo{author}{Vaughan, S.}, \bibinfo{author}{Edelson, R.},
  \bibinfo{author}{Warwick, R.}, \bibinfo{author}{Uttley, P.},
  \bibinfo{year}{2003}a.
\newblock \bibinfo{title}{On characterizing the variability properties of x-ray
  light curves from active galaxies}.
\newblock \bibinfo{journal}{Monthly Notices of the Royal Astronomical Society}
  \bibinfo{volume}{345}, \bibinfo{pages}{1271--1284}.
\bibitem[{Vaughan et~al.(2003b)Vaughan, Fabian and Nandra}]{vaughan2003x}
\bibinfo{author}{Vaughan, S.}, \bibinfo{author}{Fabian, A.},
  \bibinfo{author}{Nandra, K.}, \bibinfo{year}{2003}b.
\newblock \bibinfo{title}{X-ray continuum variability of mcg-6-30-15}.
\newblock \bibinfo{journal}{Monthly Notices of the Royal Astronomical Society}
  \bibinfo{volume}{339}, \bibinfo{pages}{1237--1255}.
\bibitem[{Welsh(1999)}]{welsh1999reliability}
\bibinfo{author}{Welsh, W.}, \bibinfo{year}{1999}.
\newblock \bibinfo{title}{On the reliability of cross-correlation function lag
  determinations in active galactic nuclei}.
\newblock \bibinfo{journal}{Publications of the Astronomical Society of the
  Pacific} \bibinfo{volume}{111}, \bibinfo{pages}{1347}.

\end{thebibliography}

\end{document}